\documentclass[default]{sn-jnl}
\usepackage{setspace}
\usepackage{ragged2e}
\usepackage{caption}
\usepackage{indentfirst}
\usepackage{physics}
\usepackage{comment}
\usepackage{fancyvrb}
\usepackage{fancyhdr}
\usepackage[export]{adjustbox}

\usepackage[position=b]{subcaption}

\newcommand{\U}{{$^{238}$U}}
\newcommand{\UF}{{$^{235}$U}}
\newcommand{\Th}{{$^{232}$Th}}

\title{Water Cherenkov muon veto for the COSINUS experiment: design and simulation optimization}

\author[1]{G. Angloher}
\author[1]{M. R. Bharadwaj}
\author[2,3]{M. Cababie}
\author[4,5]{I. Dafinei}
\author[4,6]{N. Di Marco}
\author[2,3]{L. Einfalt}
\author[5,4]{F. Ferroni}
\author[2]{S. Fichtinger}
\author[7,6]{A. Filipponi}
\author[2]{M.Friedl}
\author[2,3]{A. Fuss}
\author[8]{Z. Ge}
\author[9]{M. Heikinheimo}
\author[1]{M. N. Hughes}
\author[9]{K. Huitu}
\author[1]{M. Kellermann}
\author[2,3]{$^*$R. Maji}
\author[1]{M. Mancuso}
\author[4,6]{L. Pagnanini}
\author[1]{F. Petricca}
\author[6]{S. Pirro}
\author[1]{F. Pr\"{o}bst}
\author[7,6]{G. Profeta}
\author[6]{A. Puiu}
\author[2,3]{F. Reindl}
\author[1]{K. Sch\"{a}effner}
\author[2,3]{J. Schieck}
\author[2,3]{D. Schmiedmayer}
\author[2,3] {P. Schreiner}
\author[2,3] {C. Schwertner}
\author[1] {K. Shera}
\author[1]{M. Stahlberg}
\author[9]{A. Stendhal}
\author[4,6]{$^*$M. Stukel}
\author[6,10]{C. Tresca}
\author[2]{F. Wagner}
\author[8]{S. Yue}
\author[1]{V. Zema}
\author[8]{Y. Zhu}

\email{}

\affil[1]{Max-Planck-Institut f\"ur Physik, 80805 M\"unchen - Germany}
\affil[2]{Institut f\"ur Hochenergiephysik der \"Osterreichischen Akademie der Wissenschaften, 1050 Wien - Austria}
\affil[3]{Atominstitut, Technische Universit\"at Wien, 1020 Wien - Austria}
\affil[4]{Gran Sasso Science Institute, 67100 L'Aquila - Italy}
\affil[5]{INFN - Sezione di Roma, 00185 Roma - Italy}
\affil[6]{INFN - Laboratori Nazionali del Gran Sasso, 67010 Assergi, Italy}
\affil[7]{Dipartimento di Scienze Fisiche e Chimiche, Universit\`a degli Studi dell'Aquila, 67100 L'Aquila - Italy}
\affil[8]{SICCAS - Shanghai Institute of Ceramics, Shanghai - P.R.China 200050}
\affil[9]{Helsinki Institute of Physics, 00560 Helsinki - Finland}
\affil[10]{CNR-SPIN c/o Dipartimento di Scienze Fisiche e Chimiche, Università degli Studi dell’Aquila, 67100 L’Aquila, Italy}

\abstract{

COSINUS is a dark matter (DM) direct search experiment that uses sodium iodide (NaI) crystals as cryogenic calorimeters. Thanks to the low nuclear recoil energy threshold and event-by-event discrimination capability, COSINUS will address the long-standing DM claim made by the DAMA/LIBRA collaboration. The experiment is currently under construction at the Laboratori Nazionali del Gran Sasso, Italy, and employs a large cylindrical water tank as a passive shield to meet the required background rate. However, muon-induced neutrons can mimic a DM signal therefore requiring an active veto system, which is achieved by instrumenting the water tank with an array of photomultiplier tubes (PMTs). 
This study optimizes the number, arrangement, and trigger conditions of the PMTs as well as the size of an optically invisible region. The objective was to maximize the muon veto efficiency while minimizing the accidental trigger rate due to the ambient and instrumental background. The final configuration predicts a veto efficiency of  {99.63~$\pm$~0.16~$\%$} and 44.4~$\pm$~5.6\% in the tagging of muon events and showers of secondary particles, respectively. The active veto will reduce the cosmogenic neutron background rate to 0.11~$\pm$~0.02~cts$\cdot$kg$^{-1}$$\cdot$year$^{-1}$, corresponding to less than one background event in the region of interest for the whole COSINUS-1$\pi$ exposure of 1000 kg$\cdot$days.
}

\keywords{Dark Matter, COSINUS, Muon Veto, Dead-Layer, Cherenkov, Geant4, Monte-Carlo}

\begin{document}

\maketitle
\tableofcontents
\clearpage
\section{Introduction}\label{sec:Introduction}
Several observational evidences, including galactic rotation curves~\cite{rubin1980rotational}, gravitational lensing~\cite{ellis2010gravitational}, and cosmological modeling~\cite{bullock2017small}, indicate the presence of a non-relativistic, gravitationally interacting, and stable matter in the universe. Traditionally referred to as \textit{Dark Matter} (DM), it is expected to be responsible for 26$\%$ of the mass-energy density of our universe~\cite{aghanim2020planck} and has not been conclusively detected. A prominent, NaI(Tl) based direct detection experiment is DAMA/LIBRA~\cite{bernabei2004dark,bernabei2008dama,bernabei2015final,bernabei2022further}, which is located in the Laboratori Nazionali del Gran Sasso (LNGS) under 1400~m of rock overburden (3600 m.w.e)~\cite{ambrosio1995vertical}. For the past twenty-seven years, the DAMA collaboration has observed an annually modulating event rate whose phase and period match with the expected DM signal in an earth-bound detector~\cite{bernabei2022further}. However, under the ``standard'' DM halo assumptions~\cite{freese2013colloquium}, numerous null results of various other dark matter searches exclude the region of parameter space explored by DAMA/LIBRA~\cite{particle2020review}. These direct detection experiments employ different target materials, which makes any model-independent comparison difficult. To resolve this decade-long tension, many experiments using NaI as target material have begun taking data (ANAIS-112~\cite{amare2022dark} and COSINE-100~\cite{adhikari2022three}), are currently under construction  (COSINUS~\cite{angloher2020cosinus,angloher2016cosinus}  and SABRE~\cite{antonello2019sabre}) or are in the R$\&$D phase (ASTAROTH~\cite{Zani_2021} and PICO-LON~\cite{fushimi2016dark}).  

Uniquely, the COSINUS (Cryogenic Observatory for SIgnatures seen in Next-generation Underground Searches) experiment will operate the NaI crystals as cryogenic scintillating calorimeters~\cite{angloher2016cosinus}. The NaI detectors will be cooled to milli-kelvin temperatures, and both the scintillation light and phonon signal will be read out. The phonon signal is measured using the novel \textit{remote transition edge sensor} (remoTES) setup~\cite{angloher2023first,angloher2023particle,COSINUS2023kqd}, while scintillation light will be detected by a silicon beaker enclosing the NaI crystal and instrumented with a TES. This dual channel capability will allow for the discrimination between electromagnetic and nuclear recoil interactions on an event-by-event basis. Assuming a nuclear recoil threshold of 1~keV, a background rate similar to what is discussed in~\cite{angloher2016cosinus}, and an exposure of 1000~kg$\cdot$days, COSINUS will provide a model-independent cross-check of the DM interpretation of the DAMA/LIBRA
signal as generated by DM-nuclei scattering events~\cite{kahlhoefer2018model}.

This work details the results of a comprehensive Monte Carlo simulation to optimize the design of the active water Cherenkov muon veto for the COSINUS experiment. An optical simulation was performed to test the effect of the muon veto efficiency (probability of tagging an incident muon) with different photomultiplier tube (PMT) arrangements and trigger conditions. Additionally, a study was done on how the size of an optically invisible region in the water tank (\textit{dead layer}) would affect the overall muon veto capability and PMT trigger rate.

\section{Muon-induced backgrounds and conceptual veto design}
COSINUS is currently being built in Hall B at LNGS in Italy, with construction expected to be finished by the middle of 2024. The cooling of the detector modules is provided by a dry dilution refrigerator with the detector volume residing at its base. The cryostat is held in a stainless steel cylinder, known as the dry-well, which is contained in a larger, $\sim$7$\times$7~m, cylindrical steel tank filled with ultra-pure water that acts as a passive shield against ambient radiation. Additionally, copper shielding is placed above and around the detector volume to reduce the radiogenic background from internal components. The optimization of the passive shielding was discussed in detail in~\cite{angloher2022simulation}. Because of the event-type discrimination capability of the COSINUS experiment, neutrons are the most important background to mitigate.

 In underground laboratories, neutron flux occurs from either natural radioactivity or cosmic rays. Spontaneous fission and ($\alpha$,n) interactions from decaying isotopes of the \U, \UF\, and \Th\ chains are the processes responsible for the natural radioactivity (ambient and radiogenic neutron flux) emitted from the rock around the laboratory and detector materials. The energy spectrum of these neutrons peaks at a few MeV~\cite{kudryavtsev2008neutron} and the passive shielding discussed in~\cite{angloher2022simulation} would reduce these events to less than one neutron per year in the detector volume ($\sim$1~kg) of the COSINUS experiment.

Conversely, neutrons that are produced as a result of cosmic-ray muon interactions are much more challenging to moderate. At LNGS the cosmic muon flux was measured to be 3.4~$\times$10$^{-4}~$m$^{-2}\cdot$s$^{-1}$~\cite{bellini2012cosmic, mei2006muon}. As a result of the passive shielding of COSINUS, the dominant contribution to the neutron background will come from cosmogenics, with an expected rate of 3.5~$\pm$~0.7~cts$\cdot$kg$^{-1}$$\cdot$year$^{-1}$.  Muon-induced neutrons can be produced through four different processes: (1) negative muon capture, (2) muon-induced spallation reactions, (3) neutron production by hadrons from muon-induced hadronic cascades, and (4) neutron production by photons from muon-induced electromagnetic cascades~\cite{kudryavtsev2003simulations}. The muon-induced neutron energy spectrum can be approximated by a $1/E$ relationship up to 0.5~GeV and a $1/E^2$ relationship beyond~\cite{baxter2022snowmass2021}. The spectrum (see Fig. 5c in ~\cite{angloher2022simulation}) will extend up to a few GeV allowing these neutrons to travel through the passive shielding of the COSINUS experiment and into the detector volume. Additionally, these neutrons can be produced past the shielding, inside the setup, and transfer $>$~1~keV energies to nuclei, which will be above the threshold for the experiment. Muon-induced neutrons are typically accompanied by a large number of secondary particles. These neutrons can be detected simultaneously with the primary muon themselves or one of the secondaries created. 

\subsection{Water Cherenkov muon veto conceptual design}\label{subsec:Water_Cherenkov_Conceptual_Design}

\begin{figure}[H]
 \begin{adjustbox}{width=1.4\linewidth, right}
        \includegraphics{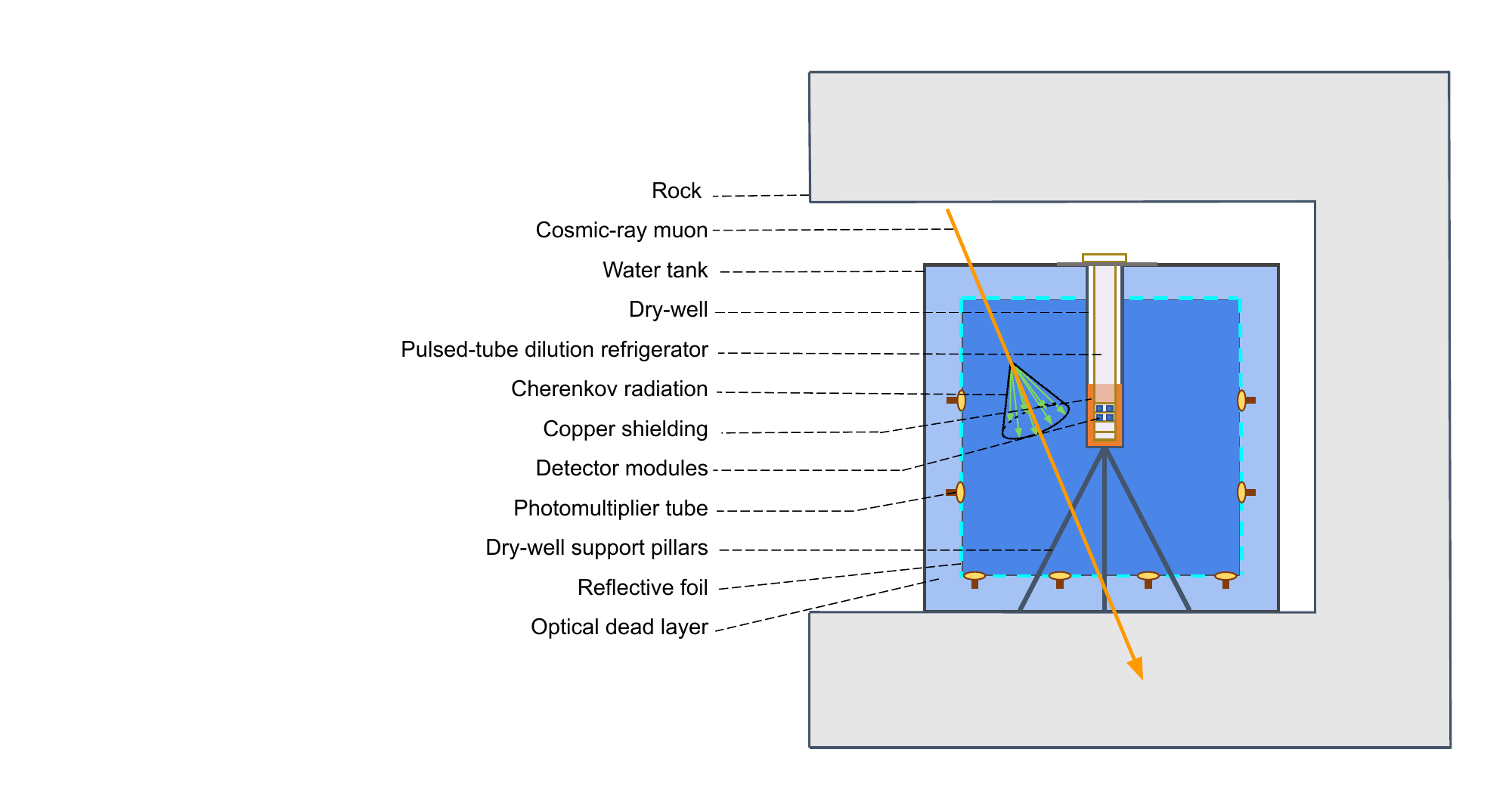}
    \end{adjustbox}
	\caption{Schematic of the simulated muon veto setup of COSINUS. The exact dimensions used in the muon veto simulation can be found in~\cite{angloher2022simulation}. }\label{Fig:Simulated_water_tank}
\end{figure}

In the COSINUS experiment the muon veto will consist of a cylindrical water tank, with PMTs mounted on the wall and bottom, and a reflector encasing the PMTs.
The PMTs will detect the Cherenkov radiation (with wavelengths in the optical range) produced from the muons or their products that enter the tank. A schematic of the veto setup can be seen in Fig.~\ref{Fig:Simulated_water_tank}.  

\begin{figure}[ht]
\begin{subfigure}[t]{0.56\textwidth}
\includegraphics[width=\textwidth]{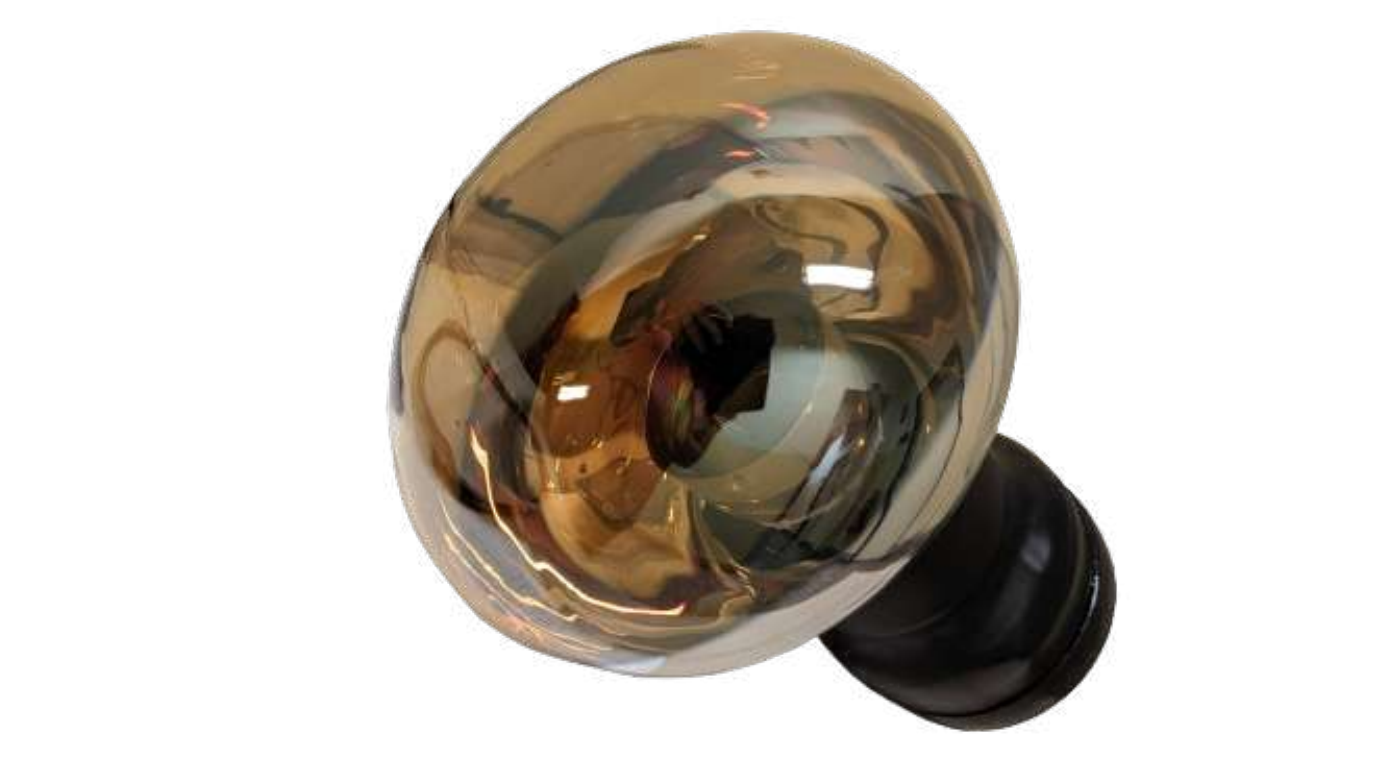}
\caption{}
\end{subfigure}
\begin{subfigure}[t]{0.43\textwidth}
\includegraphics[width=\textwidth]{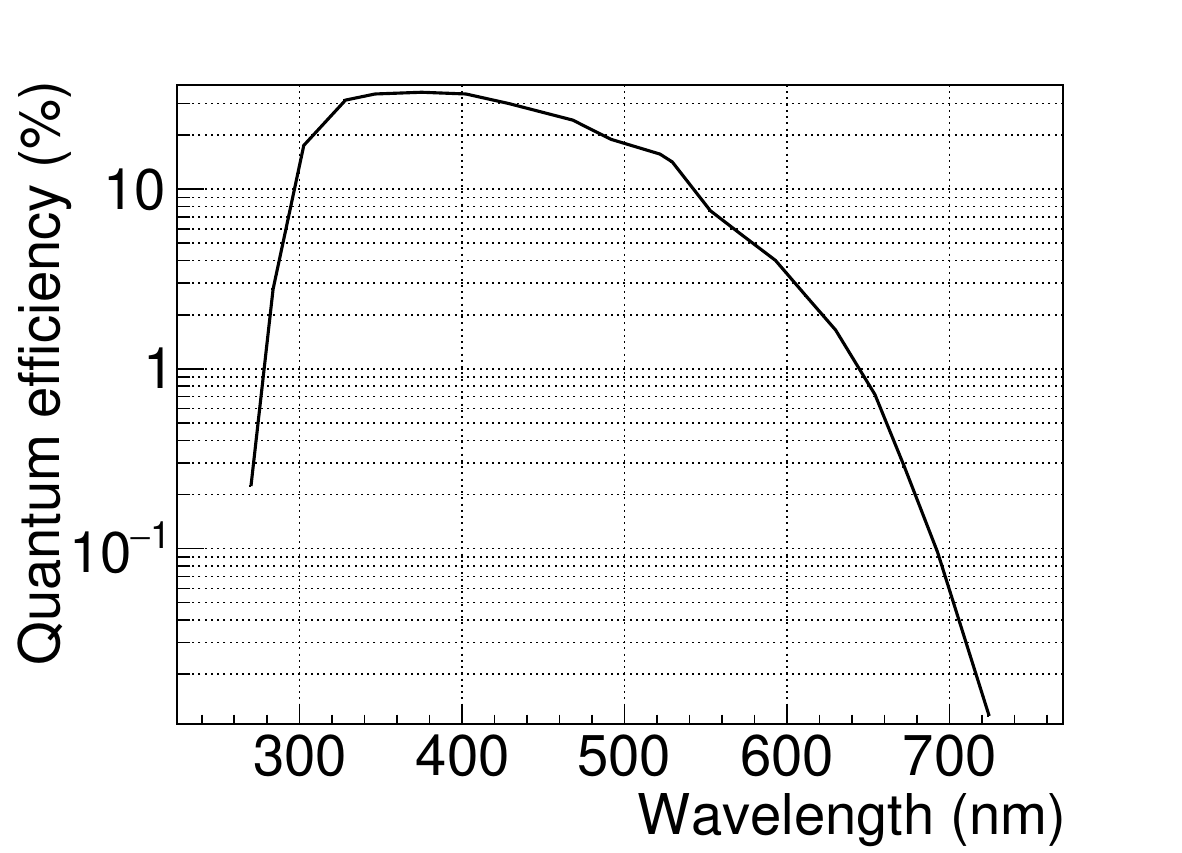}
\caption{}
\end{subfigure}
\caption{\label{Fig:Summer_run_results} (a) Hamamatsu R5912-30 photomultiplier tube, and (b) its quantum efficiency as a function of wavelength~\cite{HamamatsuPhotonics}.}\label{Fig:PMT_Quantum_Efficiency}
\end{figure}

COSINUS has selected the 8-inch diameter R5912-30 PMT from Hamamatsu~\cite{Hamamatsu_PMT_Manual}. The PMT consists of a bialkali photocathode material with borosilicate glass. A maximum of 28 PMTs will be deployed in the setup, each providing a typical gain of $\times$10$^{6}$ at an operating voltage of 1100~V. The quantum efficiency of the R5912-30 can be seen in Fig.~\ref{Fig:PMT_Quantum_Efficiency}. The efficiency of the light collection will be enhanced by the addition of a reflector, which will be placed around the interior of the water tank. The additional reflectivity will elongate the photon path increasing the probability it will be collected in a PMT.


\subsubsection{Background triggering of the PMTs}\label{subsubsec:Erroneous_PMT_Triggering}
Spurious triggers can be induced by thermionic emissions in the cathode (dark counts) and the ambient gamma background. This, in turn, can lead to random coincidences with a signal observed in the NaI detector and ultimately the rejection of a real event. Dark counts will be independent of the PMTs and depend on the specific PMT model. The nominal dark count rate for the R5912-30 is 6000~Hz, but experiments that have utilized the same PMT have measured a much lower rate of 1200~Hz~\cite{geis2018xenon1t}.

In Hall B at LNGS, the total gamma flux between 7.4-2734.2~keV was measured to be 0.23~$\gamma\cdot$cm$^{-2}\cdot$s$^{-1}$~\cite{malczewski2013gamma}. For the COSINUS experiment, this would yield a total rate through the tank of $\sim$5$\times$10$^{5}$~$\gamma\cdot$s$^{-1}$,  a factor significantly higher than the dark count rate. Although not every gamma will trigger a PMT, the effect of the ambient gammas can be reduced by creating an optically dead layer in the periphery of the tank. In~\cite{angloher2022simulation}, the gamma flux was shown to decrease by an order of magnitude for each 50~cm of water added to the radius of the tank. The idea would therefore be to split the water volume into a passive border (which will not be instrumented by the PMTs, referred to as the dead layer or optical invisible region) and an active central region. In practice, this will be achieved by placing the reflector and PMTs at a certain distance from the tank walls. The effect of the different dead layer lengths is studied in Sec.~\ref{subsec:Dead_Layer}.

\section{Description of the Monte Carlo simulation} 
For this work, a detailed Monte Carlo simulation was performed using a Geant4~\cite{GEANT4:2002zbu,Allison2006Geant,ALLISON2016186} based software toolkit called ImpCRESST~\cite{CRESST:2019oqe}, which was initially developed within the CRESST DM search experiment~\cite{abdelhameed2019first}  and later also within COSINUS. An ImpCRESST version equipped with Geant4 v10.2.3 and root v6-08-06~\cite{brun1997root} was used for this work. The ImpCRESST simulation results consist of information such as particle tracking, energy deposition, particle type, interaction mechanism, and timing information. For this particular study, the propagation of the optical photons was also recorded. 
For optimization of the muon veto design, the geometry of the set-up has been simulated. In the simulation, the experimental geometry is placed in an empty hall surrounded by $2.6$~m thick rock on the top, bottom, and one side of the water tank. The simulated geometry of the muon veto setup is shown in Fig.~\ref{Fig:Simulated_water_tank}.

\subsection{Physics models} 

The simulation's outcome relies on the processes and models that are activated and implemented through the physics list utilized. ImpCRESST employs a set of physics models adapted from the Geant4 Shielding physics list\cite{collaboration2012geant4}, which is suitable for deep shielding and neutron transport simulation. 

G4EmStandardPhysics is responsible for standard electromagnetic interaction in Geant4. In ImpCRESST, \mbox{G4EmStandardPhysics\textunderscore option4} is used, which is tailored with the most accurate models for standard and low-energy processes, with a minimum energy threshold of 100 eV (not recommended to lower below 250 eV). At the energy range of interest for DM searches, the implementation of EM physics processes happens via the various Geant4 classes. For example, Muon interactions are employed by classes such as G4MuIonisation and G4MuMultipleScattering, while interactions of $\alpha$-particles and heavier ions are governed by G4hMultipleScattering and G4ionIonisation.

G4EmExtraPhysics enables gamma-nuclear processes for simulating electromagnetic showers from high-energy muons. G4HadronElasticPhysicsHP employs high-precision models for elastic neutron scattering below 20 MeV. G4HadronPhysicsShielding manages inelastic hadron processes with HP models for neutrons below 20 MeV. G4NeutronHPThermalScattering refines elastic neutron scattering below 4 eV. G4StoppingPhysics captures charged particles at rest, including negative muons. G4IonQMDPhysics manages inelastic processes of ions above certain energy thresholds. G4ProductionCutsTable sets energy thresholds for secondary particle production, adjusted to 250 eV in ImpCRESST. G4OpticalPhysics was activated for this work, which enables the simulation of optical processes and the creation of optical photons.

\subsection{Cosmogenic muons \label{subsubsec:Cosmogenic_Muons}}
\begin{figure}[H]
 \begin{adjustbox}{width=1.1\linewidth, center}
        \includegraphics{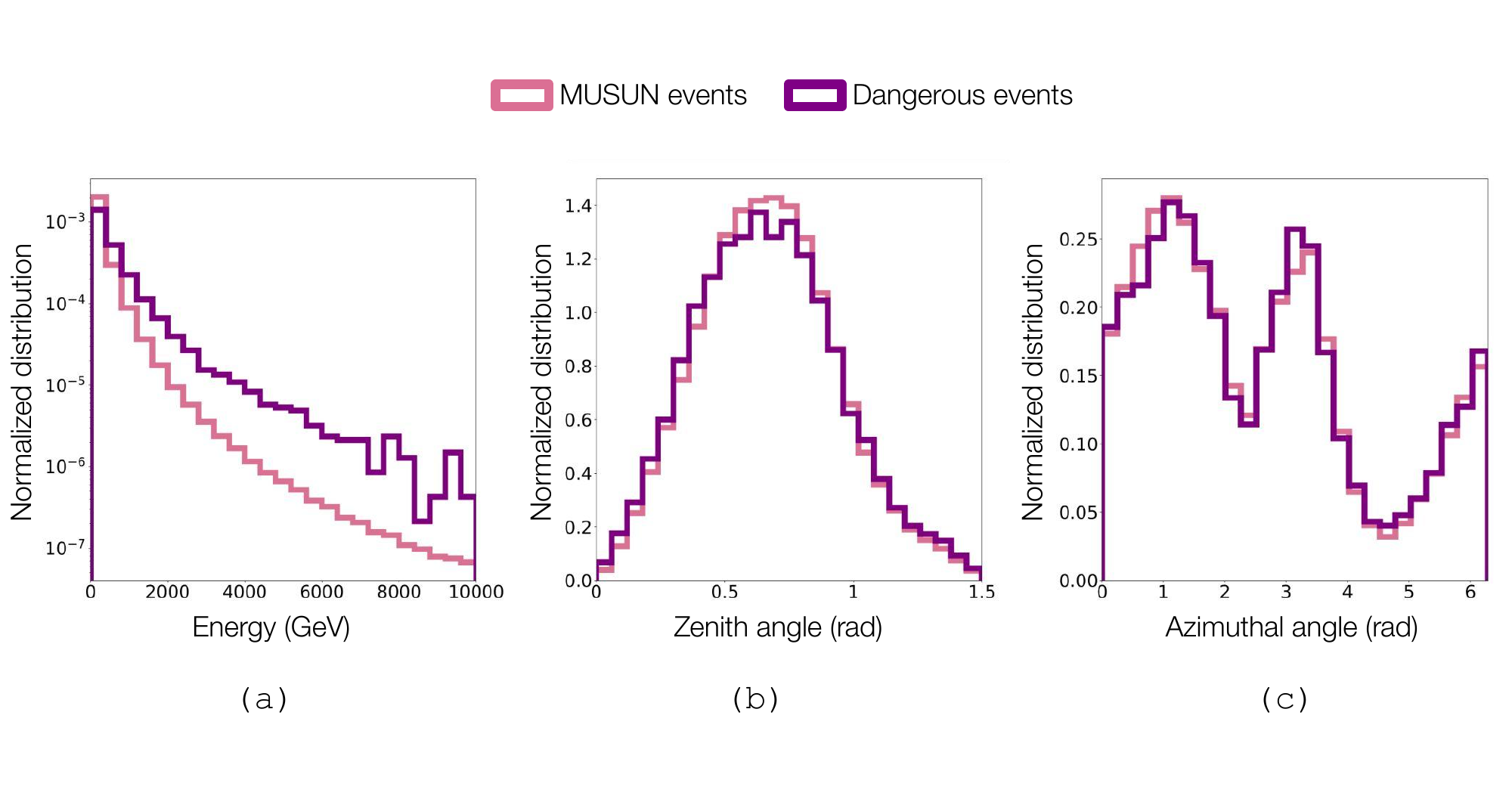}
    \end{adjustbox}
	\caption{Intial (a) energy, (b) zenith angle, and (c) azimuthal angle for the MUSUN simulated events. The normalized distributions are also shown for the subset of events that produce a neutron in the dry-well of the COSINUS experiment (i.e. the dangerous events). See text for more details.\label{Fig:MUSUN_gen_muons.pdf} }
\end{figure}

The initial position, energy, and angular distribution of cosmogenic muons underground were generated using the MUon Simulations UNderground (MUSUN)~\cite{kudryavtsev2009muon} production code. MUSUN uses the results of muon transport through rock/water to generate muons in or around underground laboratories taking into account their initial energy spectrum, angular distribution, and the overburdened rock profile. For the current work, a customized version of the code was used that incorporates the complex mountain landscape of the LNGS site~\cite{kudryavtsev2008neutron}. 
\\
Drawing from the MUSUN-generated events, originating from a 12$\times$12$\times$13~m$^{3}$ cuboid surface, 30 million muons were simulated through the COSINUS geometry. This corresponds to an exposure of $\sim$13 years which is approximately 4 times the planned measurement length of COSINUS. If a muon produces a neutron that enters the dry-well of the experiment this event is classified as \textit{dangerous}. Out of 30 million muons, 11682 were tagged as dangerous with their initial energy and angular distribution shown in Fig.~\ref{Fig:MUSUN_gen_muons.pdf}. These dangerous muons are further classified into two categories: 1) \textbf{muon events} where the muon passes directly through the water tank itself generating Cherenkov radiation and 2) \textbf{shower events} where the muon misses the tank, interacts in the rock or cryostat components creating secondary particles that produce the Cherenkov radiation. The simulations show that shower events will make up $\sim$5$\%$ of all dangerous events to reach the cryostat. The optical simulations were conducted exclusively on dangerous events to evaluate the impact of the different muon veto configurations on reducing the cosmogenic neutron background.

\subsection{Ambient gamma background}\label{subsubsec:Ambient_Gamma_Background}

As discussed in Sec.~\ref{subsubsec:Erroneous_PMT_Triggering}, ambient gammas contribute significantly to the overall PMT trigger rate. This necessitates an optical simulation in order to evaluate the best possible configuration to reduce the ambient gamma trigger rate. For this work, 20 million ambient gammas are simulated, corresponding to an exposure of 40~s. The sampled energy distribution is shown in Table~\ref{tab:Ambient_Gamma_Flux} which is measured in Hall B of LNGS~\cite{malczewski2013gamma}. Ambient gammas are generated from the internal surface of a cylindrical volume, positioned at a distance of 1~mm outside the water tank, with gamma rays directed inward towards the tank. Sec.~\ref{subsubsec:Ambient_Gamma_Background_Rate} presents a detailed description of the results of the simulation. 

\begin{table}[ht]
\centering
\begin{tabular}{cc}\hline
Energy Region (keV) & Flux (~$\gamma\cdot$cm$^{-2}\cdot$s$^{-1}$) \\\hline
7.4 - 249.8 & 0.137 \\
250.2 - 500.4 & 4.24$\times$10$^{-2}$ \\
500.8 - 1005.2 & 2.99$\times$10$^{-2}$ \\
1005.6 - 1555.8 & 1.46$\times$10$^{-2}$ \\
1556.2 - 2055.8 & 3.50$\times$10$^{-3}$ \\
2056.2 - 2734.2 & 2.02$\times$10$^{-3}$ \\\hline
\end{tabular}
\caption{\label{tab:Ambient_Gamma_Flux} Ambient gamma flux as a function of energy in Hall B of LNGS. Adopted from~\cite{malczewski2013gamma}.}
\end{table}

\section{Simulation results}\label{sec:Sim_Results}
The following chapter details the results of the design study to optimize the tagging efficiency of a water Cherenkov muon veto. In Sec.~\ref{subsec:Muon_Illumination_Map}, the optical photon maps along the tank walls from the incident cosmogenic muons are presented. That information is then used to test the efficiency of different PMT arrangements, reflector type, and trigger conditions in Sec.~\ref{subsec:PMT_Arrangement} and~\ref{subsec:Experimental_Design_Configurations}. To reduce the ambient background trigger rate various optical dead layers were studied in Sec.~\ref{subsec:Dead_Layer}. The information found in the following sections was then used to recommend the final veto setup, found in Sect.~\ref{subsec:Final_configuration}.

\subsection{Cherenkov illumination of the water tank}\label{subsec:Muon_Illumination_Map}
\begin{figure*}[ht]
\begin{subfigure}[t]{0.495\textwidth}
\includegraphics[width=\textwidth]{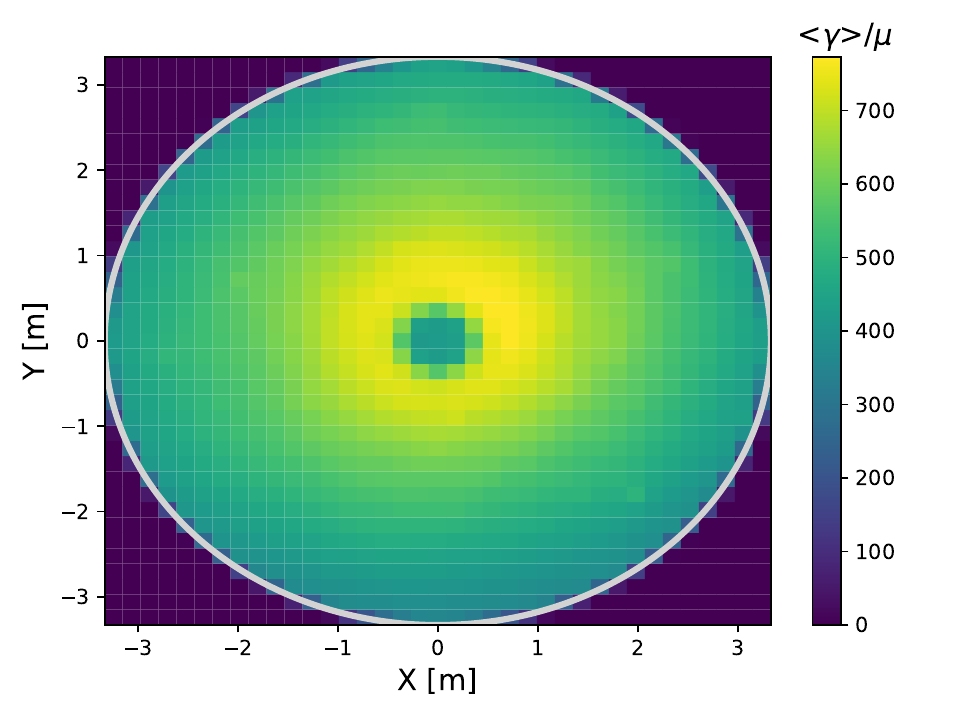}
\caption{}
\end{subfigure}
\begin{subfigure}[t]{0.495\textwidth}
\includegraphics[width=\textwidth]{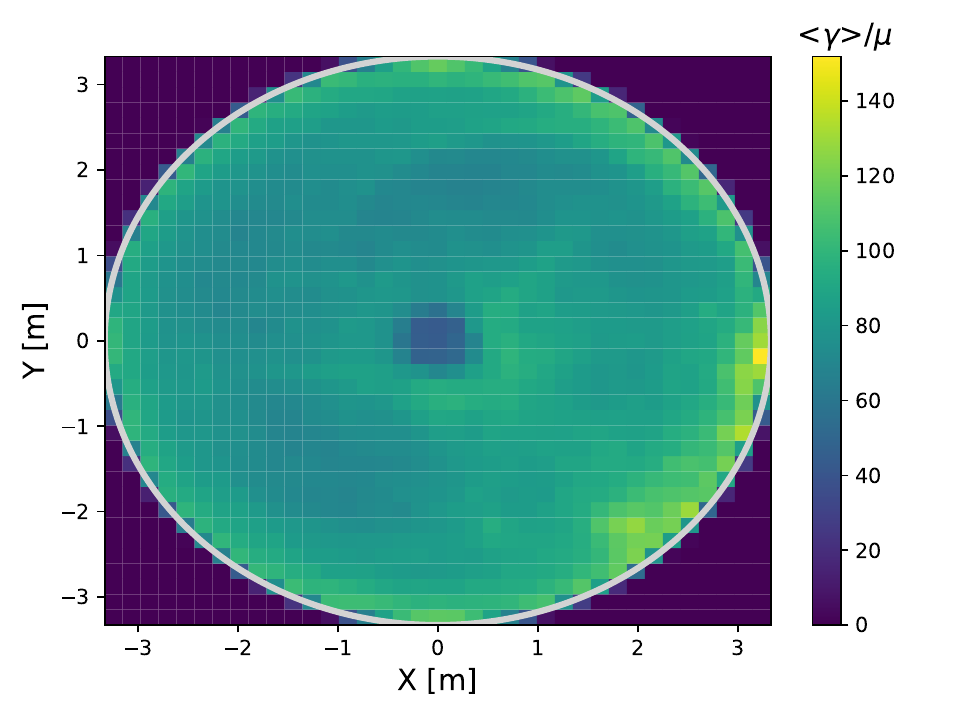}
\caption{}
\end{subfigure}
\caption{\label{Fig:Photon_Map_Bottom} Photon illumination maps on the bottom of the water tank generated by muon(a) and shower(b) events. Binning corresponds to the surface area of an 8-inch diameter PMT. The Z-axis is the average number of photons that interact in a specific bin area on the bottom surface.}
\end{figure*}

\begin{figure*}[ht]
\begin{subfigure}[t]{0.49\textwidth}
\includegraphics[width=\textwidth]{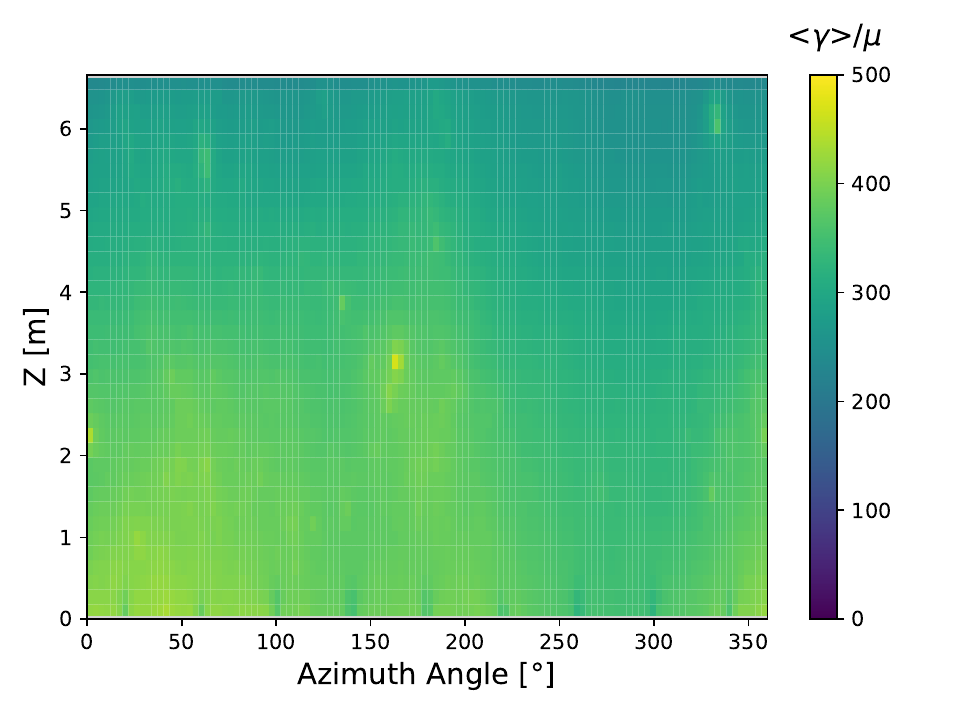}
\caption{}
\end{subfigure}
\begin{subfigure}[t]{0.49\textwidth}
\includegraphics[width=\textwidth]{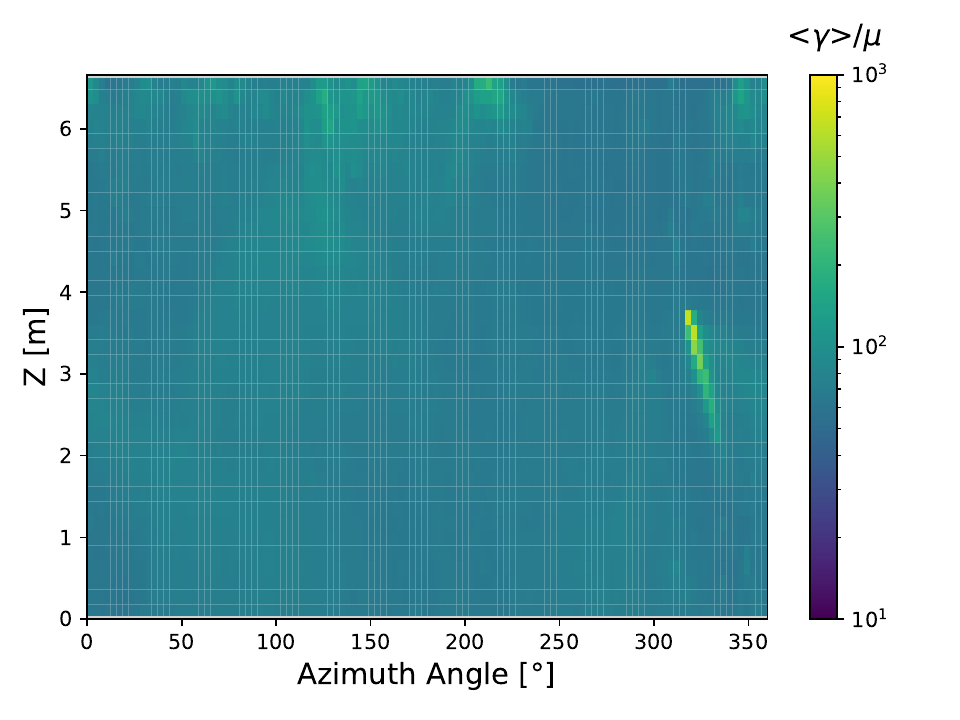}
\caption{}
\end{subfigure}
\caption{\label{Fig:Photon_Map_Wall}Muon (a) and shower (b) event photon illumination map of the wall of the water tank. Binning corresponds to the surface area of the 8-inch diameter PMT. The Z-axis is the average number of photons that interact in a specific bin area on the wall surface.}
\end{figure*}

To study the veto efficiency with different PMT configurations, the interaction points of the Cherenkov photons, produced from muon (or shower) events, were mapped along the surface of the COSINUS water tank. This is shown for both the bottom (Fig.~\ref{Fig:Photon_Map_Bottom}) and the wall (Fig.~\ref{Fig:Photon_Map_Wall}) of the tank. In the simulation, the tank is layered with a thin polyethylene material with a reflectivity of 95$\%$. The contrast between the photon maps of muon and shower events is due to the topology of the event and how the particles are absorbed/transported through the water tank. Most muon events will traverse through the water tank, producing copious amounts of Cherenkov radiation as they pass. This will tend to give a more uniform spread where any unique, visible features in the illumination map will be due to the directionality of the muons and the LNGS mountain rock profile (taken into account by the MUSUN simulation). The products of shower events, however, will get absorbed closer to the surface of the water tank, inducing a larger collection of photons around the border (see Fig.~\ref{Fig:Photon_Map_Bottom}b). Occasionally, shower products can produce intense, localized traces, such as the ones seen in Fig.~\ref{Fig:Photon_Map_Bottom}b and~\ref{Fig:Photon_Map_Wall}b. 
The inside of the COSINUS water tank has very little material and the main shadowing is done by the dry-well containing the cryostat. This produces the dull circle seen in the middle of Fig.~\ref{Fig:Photon_Map_Bottom}a and b.
Qualitatively these photon maps provide a guide on how the PMTs should be placed. A majority of the PMTs should be on the bottom of the water tank as this corresponds to a higher concentration of photons. Additionally, the bottom PMTs can be arranged in two concentric circles, one with a smaller diameter ($\sim$1.5~m) for the muon events and one with a diameter closer to the border region for shower events. Determining the position of the wall PMTs from the photon illumination map is more ambiguous and specific PMT arrangements are discussed in Sec.~\ref{subsec:PMT_Arrangement}, along with different trigger conditions. 

\subsection{Photomultiplier arrangement and trigger conditions}\label{subsec:PMT_Arrangement}

\begin{figure*}[ht]
\begin{subfigure}[t]{0.495\textwidth}
\includegraphics[width=\textwidth]{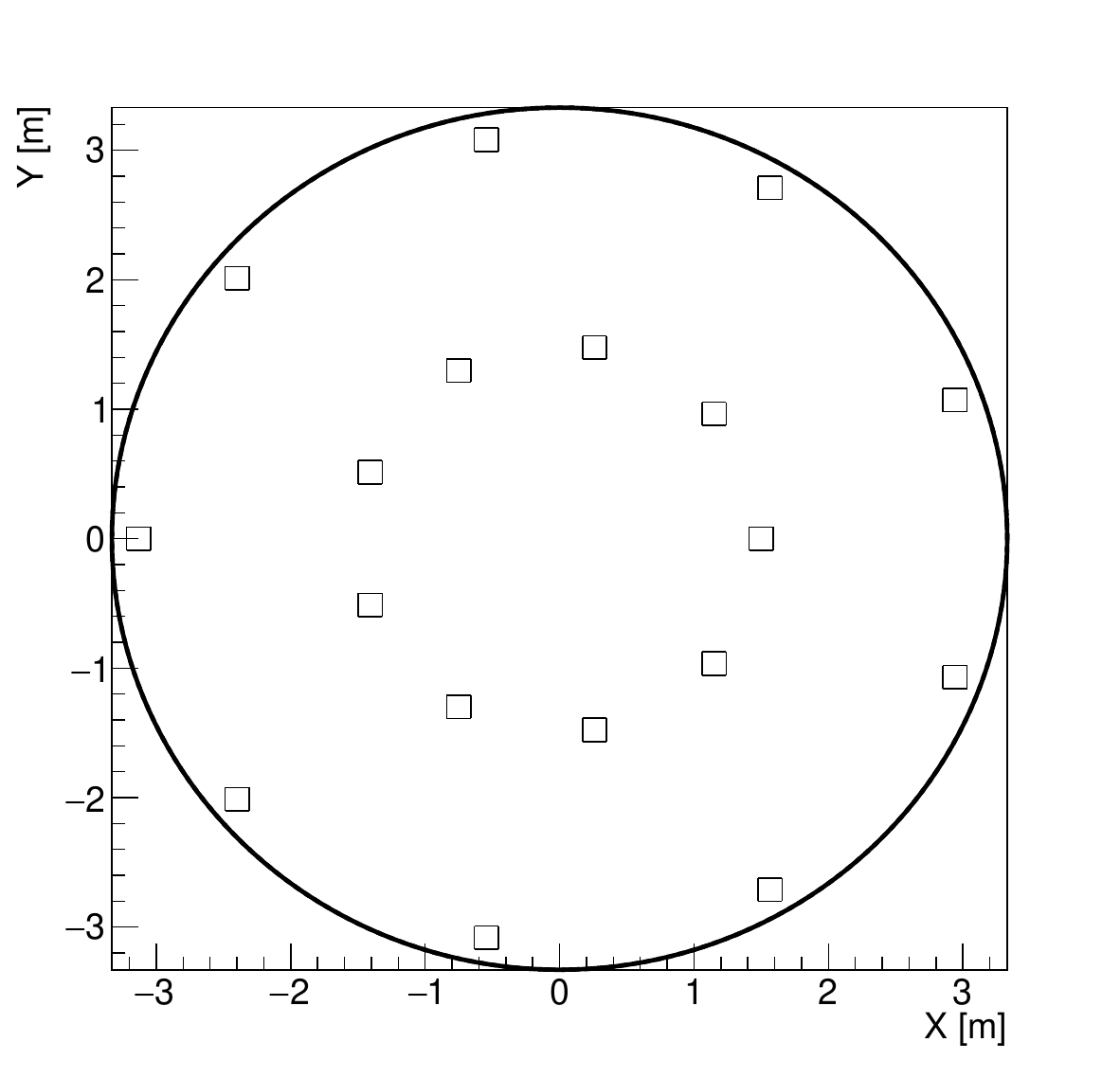}
\caption{}
\end{subfigure}
\begin{subfigure}[t]{0.495\textwidth}
\includegraphics[width=\textwidth]{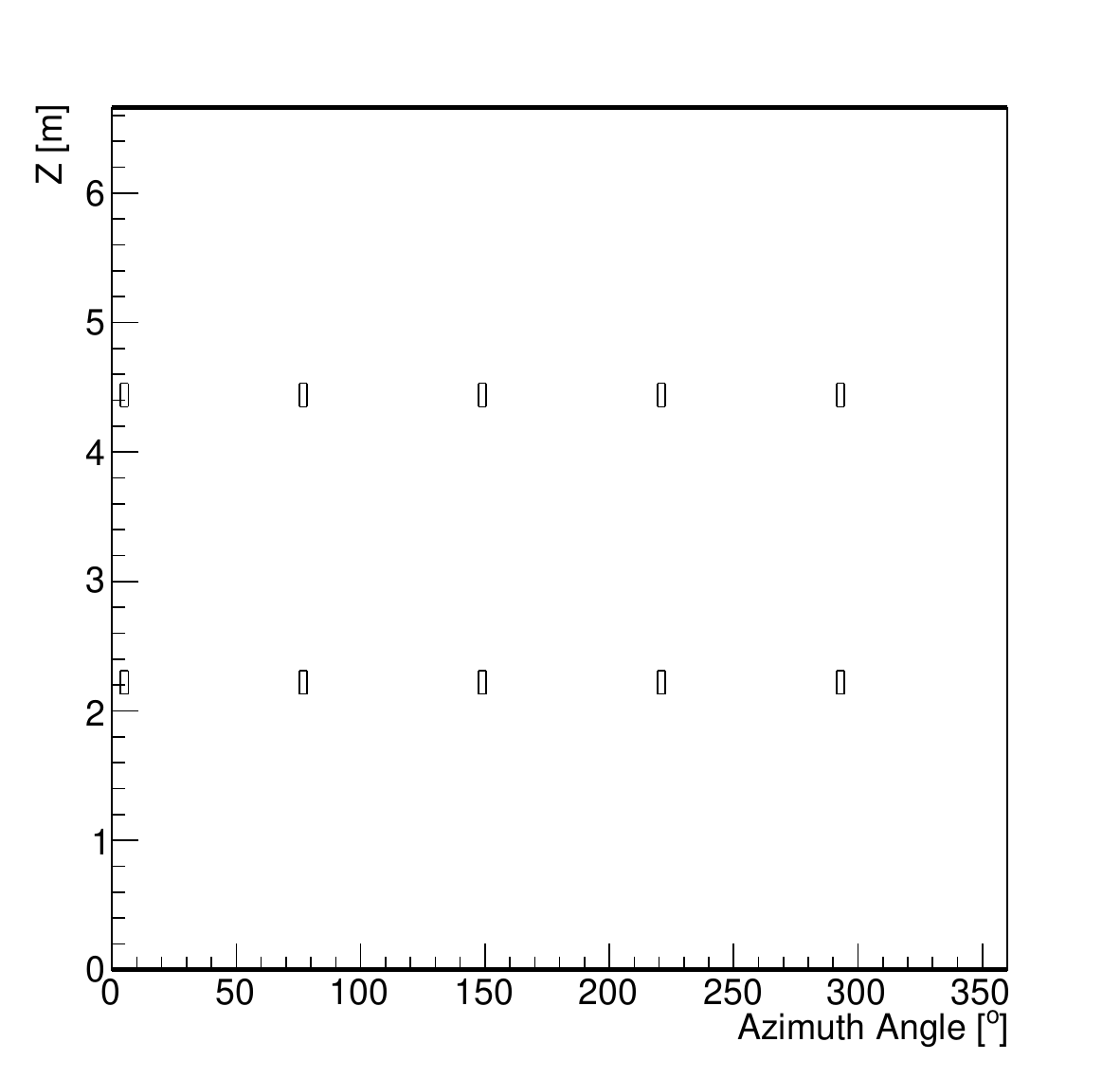}
\caption{}
\end{subfigure}
\caption{\label{Fig:PMT_Arrangement} Example 28 PMT arrangement along the bottom (a) and wall (b) of the water tank.}
\end{figure*}

The PMTs employed in the COSINUS muon veto, will be placed both on the bottom and the wall of the water tank. Following the qualitative arguments presented in Sec.~\ref{subsec:Muon_Illumination_Map} a standard arrangement can be seen in Fig.~\ref{Fig:PMT_Arrangement}. PMTs are represented along the surface as a $\sim$18$\times$18~cm$^{2}$ square which gives them an equivalent surface area to the 8-inch diameter PMT. A collection efficiency of 90$\%$ is assumed based on the effective area of the R5912-30 PMT. The quantum efficiency value is taken from Fig.~\ref{Fig:PMT_Quantum_Efficiency}. When a photon interacts with the PMT surface it is subjected to a quantum efficiency and collection efficiency test. Passing both of these, the photon counts as a \textit{photoelectron hit} for that PMT. The PMT is then triggered if the total number of photoelectron hits for a given event equals or exceeds a predetermined threshold. The final veto condition is then set by requiring an L-fold triggering of PMTs for a given event.  

\begin{figure*}[ht]
\centering
\begin{subfigure}[t]{0.6\textwidth}
\includegraphics[width=\textwidth]{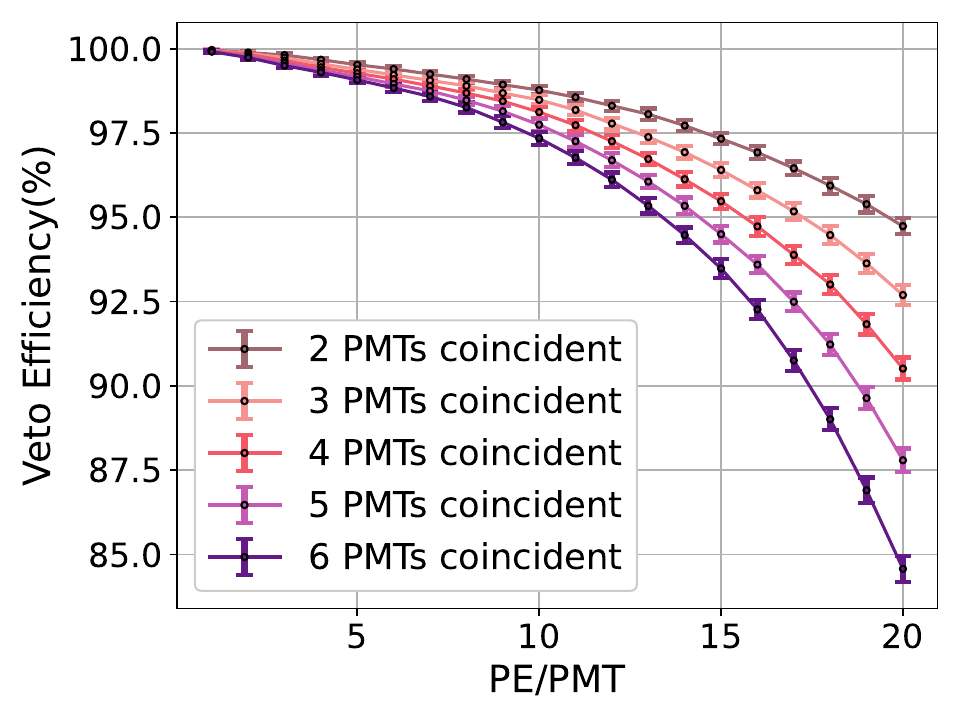}
\caption{}
\end{subfigure}

\begin{subfigure}[t]{0.6\textwidth}
\includegraphics[width=\textwidth]{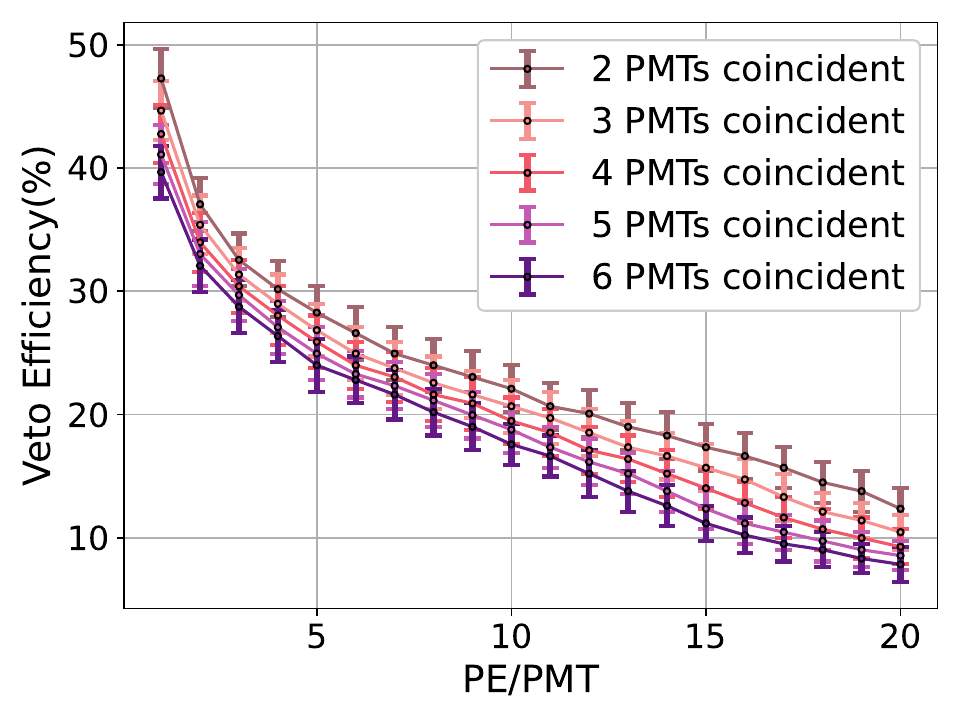}
\caption{}
\end{subfigure}
\caption{\label{Fig:Veto_Efficiency} Muon (a) and shower (b) veto efficiency as a function of photoelectron threshold (PE) per PMT. For a given photoelectron trigger, the shower veto efficiency does not change by more than 10\% over the required number of PMTs in coincidence. The muon veto efficiency variance, however, depends strongly on the number of photoelectrons to trigger.The uncertainty shown was determine using the bootstrap method~\cite{davison1997bootstrap}.}
\end{figure*}

Using the dangerous muons from Sec.~\ref{subsubsec:Cosmogenic_Muons} as the initial particle source and the PMT arrangement shown in Fig.~\ref{Fig:PMT_Arrangement}, the veto efficiency is evaluated at different trigger conditions. This is shown in Fig.~\ref{Fig:Veto_Efficiency} for both muon and shower events. In general, Fig.~\ref{Fig:Veto_Efficiency} shows that both a high photoelectric threshold or number of PMTs in coincidence will result in a decrease of vetoing power. The large amount of Cherenkov light produced during a muon event will still result in a large veto efficiency of at least 85$\%$, even with the strictest of triggering requirements. Conversely, due to the lack of Cherenkov photons, shower events experience a large drop-off in vetoing power with an increased photoelectron threshold. This necessitates a low photoelectron number threshold in order to be as efficient as possible in tagging the shower events, and even then only $\sim$40$\%$ of shower events will be vetoed.

\begin{figure}[ht]
  \centering
  \includegraphics[width=0.8\linewidth]{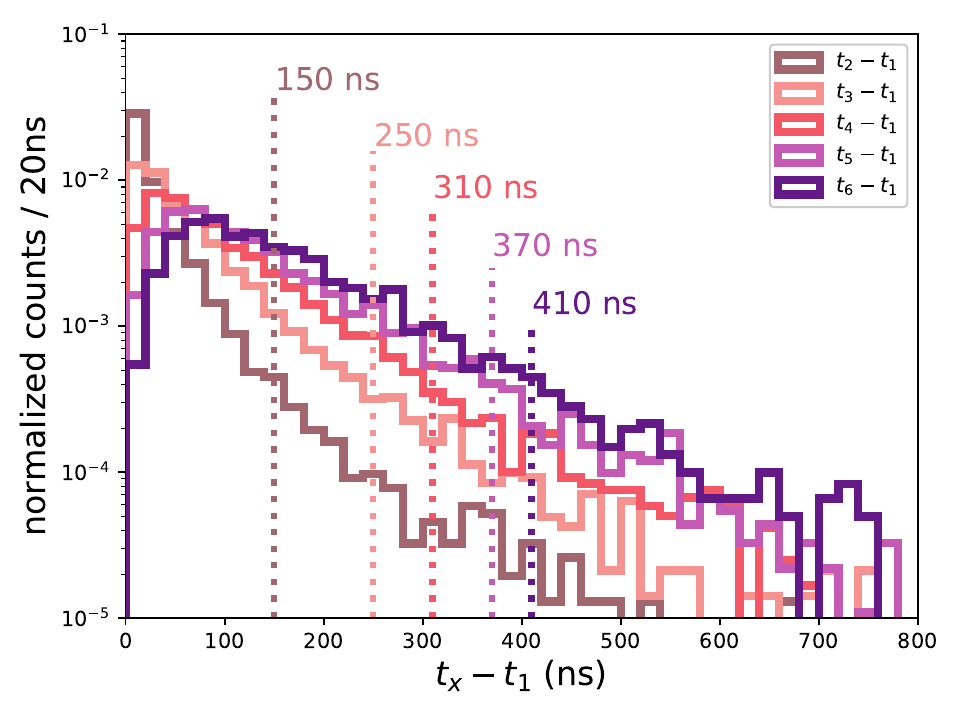}
	\caption{\label{Fig:Muon_Time_Distribution.pdf} Distribution of the time difference between photons arriving at the first and final triggered PMT. A various number of PMT coincidence requirements are shown. With the 2-fold PMT coincidence defined by $t_2$, 3-fold by $t_3$, etc. The 95$\%$ event integral for each PMT fold coincidence is denoted by the corresponding dotted lines.}
\end{figure}

The other parameter to be investigated is the time width of the coincidence window. The time difference between the triggering of the first and last PMT is shown in Fig.~\ref{Fig:Muon_Time_Distribution.pdf}. These distributions contain both muon and shower events, and it is not expected that the event type will have an effect on the overall shape. By integrating over the spectra, the required coincidence window can be determined in order to successfully tag the event. For a 6-fold PMT coincidence and a photoelectric threshold per PMT of 1, 95$\%$ of events will be tagged with a coincidence window of 410~ns.

\subsection{Experimental design considerations}\label{subsec:Experimental_Design_Configurations}

\begin{figure}[ht]
  \centering
  \includegraphics[width=0.8\linewidth]{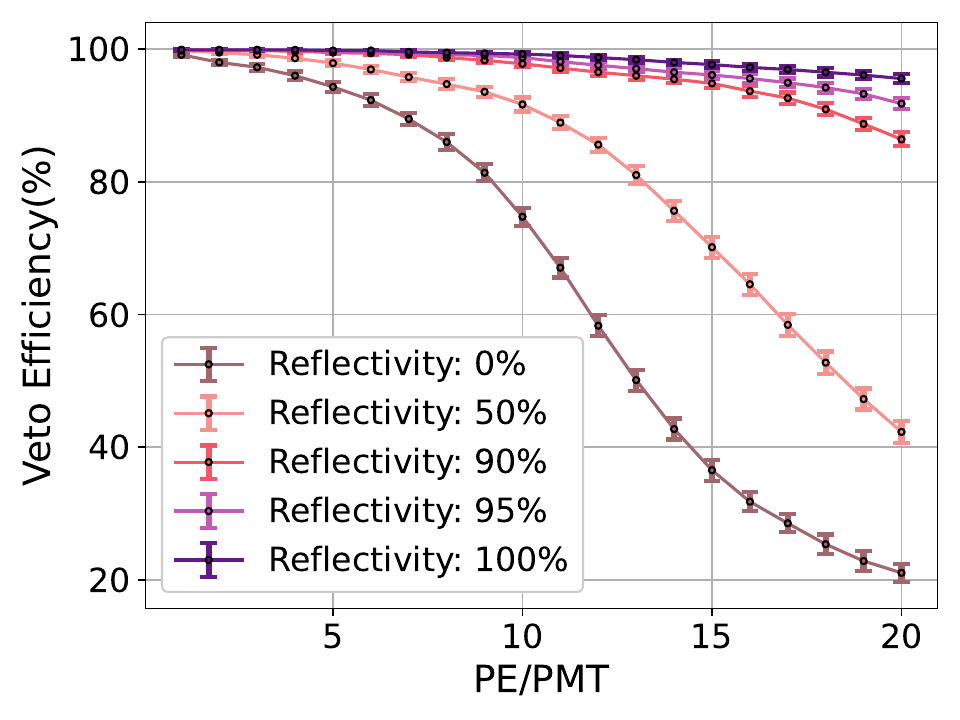}
	\caption{\label{Fig:Reflectivity_Muon_Veto.pdf} Effect of the reflectivity of the reflector on the muon veto efficiency. The uncertainty shown was determine using the bootstrap method~\cite{davison1997bootstrap}.}
\end{figure}

The reflector can play a large role in the strength of the veto, as the photon survival probability and thus PMT hit rate will increase with reflectivity. The muon event veto efficiency was studied for a reflectivity of 0, 50, 90, 95, and 100$\%$ in Fig.~\ref{Fig:Reflectivity_Muon_Veto.pdf}. This figure shows that at a low photoelectron threshold, the reflectivity will have a minimal effect on veto efficiency and the veto power will only severely deteriorate with a high threshold. The surface type (specular vs. diffuse) of the reflector was also studied. It was found that, in all trigger conditions, a diffuse reflector resulted in a drop of 5$\%$ efficiency in vetoing shower events, while the muon event veto efficiency remained unchanged at low photoelectron thresholds. As such, a specular reflector is recommended. 

An assortment of PMT arrangements was tested in order to search for an optimal configuration that maximizes the muon and shower event veto efficiency. This included varying the parameters:

\begin{itemize}
    \setlength{\itemindent}{1cm} 
    \item inner ring radius along the bottom of the water tank
    \item outer ring radius along the bottom of the water tank
    \item number of rows for the wall PMTs
    \item total number of PMTs along the bottom of the water tank
    \item total number of PMTs along the top of the water tank
    \item vertical position shifting of the wall PMT rows
\end{itemize}

However, the only factor that significantly ($>$ 1$\%$ change) altered the efficiency was the total number of PMTs along the bottom or the wall of the tank. A linear relationship was found between the veto efficiency and the number of PMTs but, as COSINUS is restricted to a total number of 28 PMTs (to balance efficiency and cost) this cannot be exploited further. Therefore, based on construction restrictions, the configuration shown in Fig.~\ref{Fig:PMT_Arrangement} (or some small variation to account for the dead layer) will be the one used for the rest of this study.

\subsection{Optimization of an optical dead layer}\label{subsec:Dead_Layer}

One design goal of the water Cherenkov veto is to maximize the muon tagging efficiency, while minimizing the PMT trigger rate from non-muon sources. Based on the discussion in Sec.~\ref{subsec:PMT_Arrangement} and~\ref{subsec:Experimental_Design_Configurations}  there are three primary parameters that can be tuned in the optimization: 1) the size of the optical dead layer, 2) number of PMTs required to trigger to classify an event and 3) the PE-threshold. The following sections discuss the optimization of these for the COSINUS water Cherenkov muon veto.

\subsubsection{Ambient gamma background rate}\label{subsubsec:Ambient_Gamma_Background_Rate}
As discussed in Sec.~\ref{subsubsec:Erroneous_PMT_Triggering} the PMTs can be triggered by other sources than cosmic ray muons. Outside of dark counts, the primary background will be from ambient gammas. Following the procedure outlined in Sec.~\ref{subsubsec:Ambient_Gamma_Background} an optical simulation of the ambient gamma background was performed. The reflectivity was chosen to be 95$\%$ and the PMTs were arrayed along the wall and bottom of the water tank, as shown in Fig.~\ref{Fig:PMT_Arrangement}. Using the same triggering method as described in Sec.~\ref{subsec:PMT_Arrangement} and a single photoelectron threshold it was found that the rate of triggering two PMTs, caused by the ambient gammas, was 359~$\pm$~3 Hz. The probability of having a spurious coincidence over a specified time window $\Delta T$ with a given background rate $R_{B}$ is given by 
\begin{equation}\label{Eqn:Spurious_Coinc_Prob}
    p(\Delta T) = \int^{\Delta T}_{0} R_B e^{-R_B T} dT = 1-e^{-R_B \Delta T}.
\end{equation}
A typical time window around a cryogenic event is $\sim$10~ms~\cite{angloher2023first}. In order to have a $<$1$\%$ chance of a spurious coincidence the ambient gamma trigger rate, employing Eq.~\eqref{Eqn:Spurious_Coinc_Prob}, would need to be $<$1~Hz.

\begin{figure}[ht]
  \centering
  \includegraphics[width=0.8\linewidth]{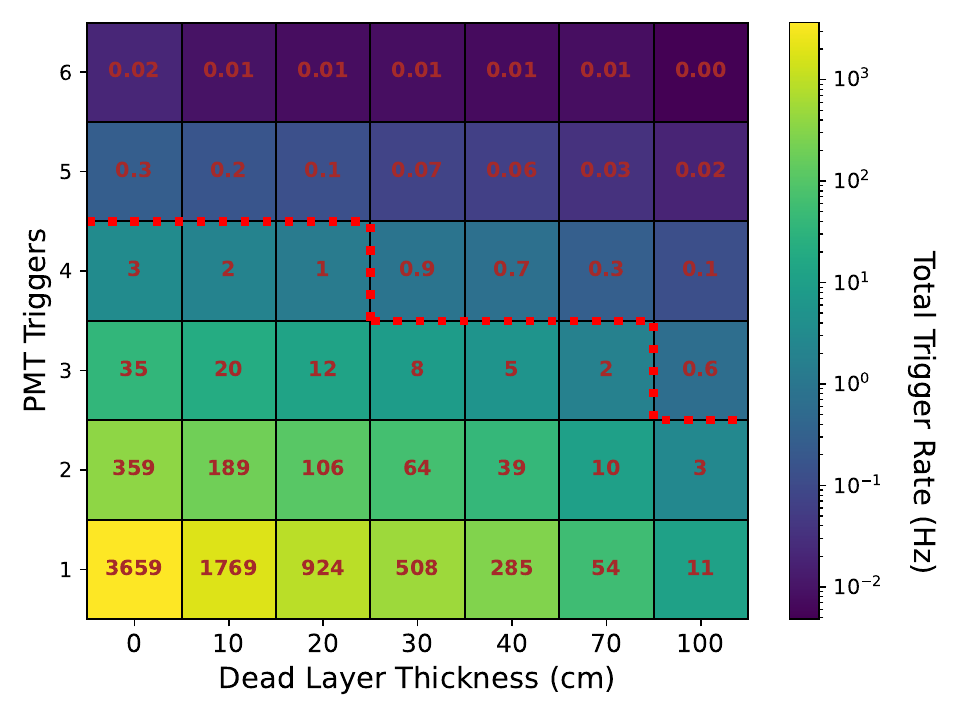}\caption{Rate of PMT triggers caused by the ambient gammas when considering different dead layers. Above the dashed red line denotes configurations in which the trigger rate is less than 1~Hz. The rates shown have been corrected with respect to the raw simulated rates as described in~App.~\ref{appendix:subsection:true_L_fold_coincidence}}\label{Fig:Ambient_Gamma_Rate_2D.pdf}
\end{figure}

The ambient gamma trigger rate can be reduced by creating an optical dead layer (described in Sec.~\ref{subsubsec:Erroneous_PMT_Triggering}) or by requiring a higher-fold PMT coincidence. An optical dead layer will decrease the overall ambient gamma flux present in the active volume of the water tank, reducing the total trigger rate of the PMTs. Additionally, due to the small amount of light produced from the ($\sim$~100~keV) $\gamma$-rays, requiring a higher PMT coincidence will also reduce the overall trigger rate. It is possible to reduce the ambient gamma trigger rate with a larger photoelectron threshold, but, as shown in Fig.~\ref{Fig:Veto_Efficiency}, this will unnecessarily reduce the effective muon and shower veto efficiency. For six different dead layer thicknesses (0, 10, 20, 30, 40, 70, and 100~cm), the ambient gamma background was simulated using the same reflector and PMT arrangement as discussed above. The results can be seen in Fig.~\ref{Fig:Ambient_Gamma_Rate_2D.pdf}, where the veto configurations above the dashed-red line are those that reduce the ambient gamma trigger rate below 1~Hz. Fig.~\ref{Fig:Ambient_Gamma_Rate_2D.pdf} shows that for each additional PMT that is required to be in coincidence, the ambient gamma trigger rate will drop by approximately an order of magnitude. Conversely, for every 10~cm of dead layer, the rate will decrease by a factor of two. Additional dead layer increases are less effective above 40~cm.

\begin{figure*}[ht]
\begin{subfigure}[t]{0.495\textwidth}
\includegraphics[width=\textwidth]{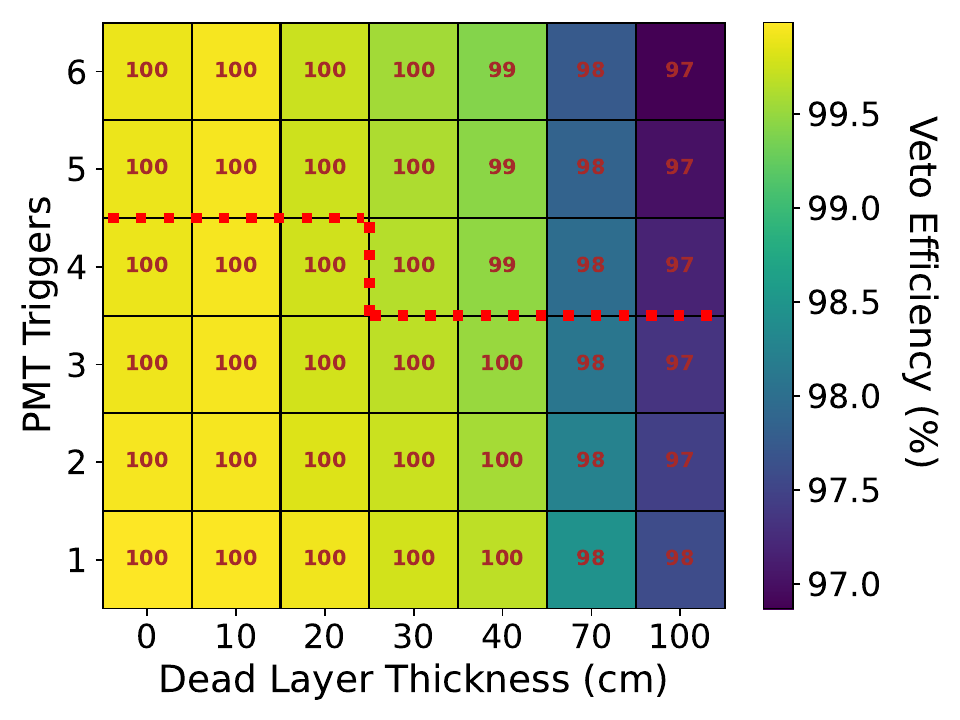} 
\caption{}
\end{subfigure}
\begin{subfigure}[t]{0.495\textwidth}
\includegraphics[width=\textwidth]{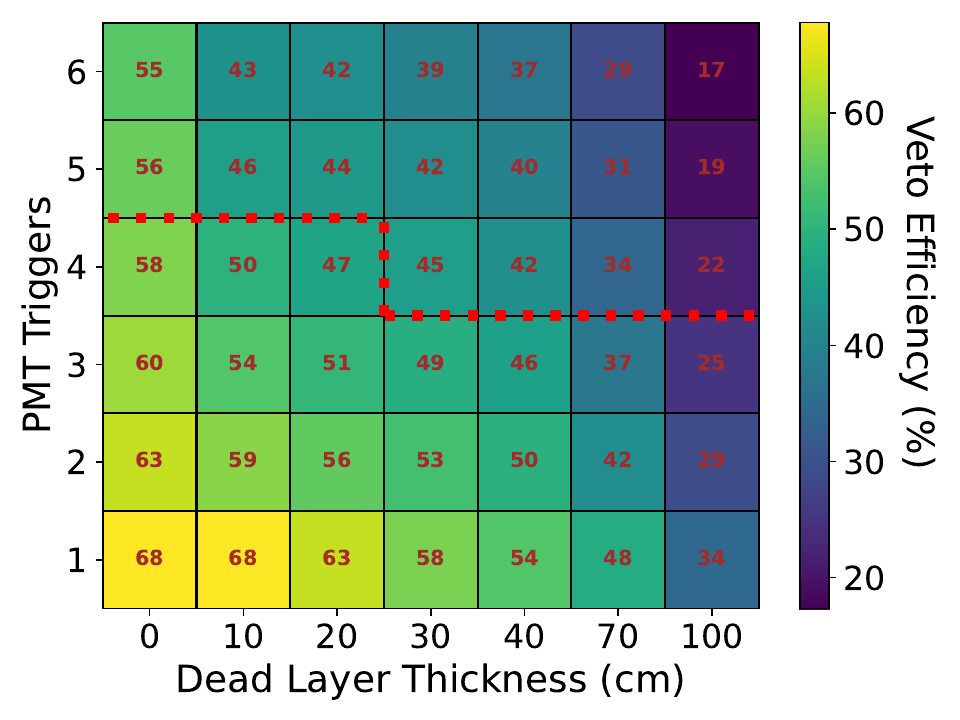}
\caption{}
\end{subfigure}
\caption{\label{Fig:Dead_Layer_Comparison} Muon (a) and shower (b) event veto efficiency under the effect of different dead layers. Multiple L-fold PMT triggering is evaluated. The same dashed red line from Fig.~\ref{Fig:Ambient_Gamma_Rate_2D.pdf} is overlayed, indicating which configurations would result in an ambient background less than 1~Hz.}
\end{figure*}

Unfortunately, it is not possible to increase the dead layer or PMT coincidence requirement without reducing the overall veto efficiency. Following the same procedure outlined in Sec.~\ref{subsec:PMT_Arrangement} the veto efficiency was evaluated for different dead layer thicknesses and PMT trigger conditions, shown in Fig.~\ref{Fig:Dead_Layer_Comparison}. Again a single photoelectron threshold was used to maximize the efficiency. The results clearly show that increasing the dead layer thickness results in a decrease in overall veto efficiency. The overall reduction in efficiency is vastly different when comparing muon and shower events. Muon events, which generate a large amount of light, are not affected by higher PMT coincidence requirements and only lose a few percent in efficiency with a large, 100~cm dead layer. Shower events will see a few percent reductions in efficiency for each increase in PMT coincidence requirement but will be dramatically reduced by an increase in the dead layer. When five PMTs are required to be in coincidence there is a drop of 36$\%$ (from 56$\%$ to 19$\%$) when comparing 0 to 100~cm dead layers. This drop is due to the fact, discussed in Sec.~\ref{subsec:Muon_Illumination_Map}, that the distribution of light from the shower event falls primarily near the tank's surface. From Fig.~\ref{Fig:Ambient_Gamma_Rate_2D.pdf} and ~\ref{Fig:Dead_Layer_Comparison} it is clear that a higher fold PMT requirement is more effective than an increased dead layer at reducing the false trigger rate while maintaining the overall veto power.

Taking into consideration that the ambient gamma trigger rate must be below 1~Hz the optimal configuration that maximizes the veto efficiency was found to be: a single photoelectron threshold, a 5-fold PMT coincidence requirement, and no dead layer. This gives a muon event veto efficiency of 99.89~$\pm$~0.05~\% and a shower event veto efficiency of 56.19~$\pm$~4.8~\%.

It should be noted that the systematic uncertainties in the muon propagation and interaction mechanisms in GEANT4 and MUSUN are not included in this paper as that is beyond the scope of this work. 

\subsubsection{Inclusion of thermionic emissions in the total background rate}\label{subsubsec:Total_Background_Rate}

\begin{figure}
\begin{subfigure}{0.495\textwidth}
    \centering
    \includegraphics[width=\textwidth]{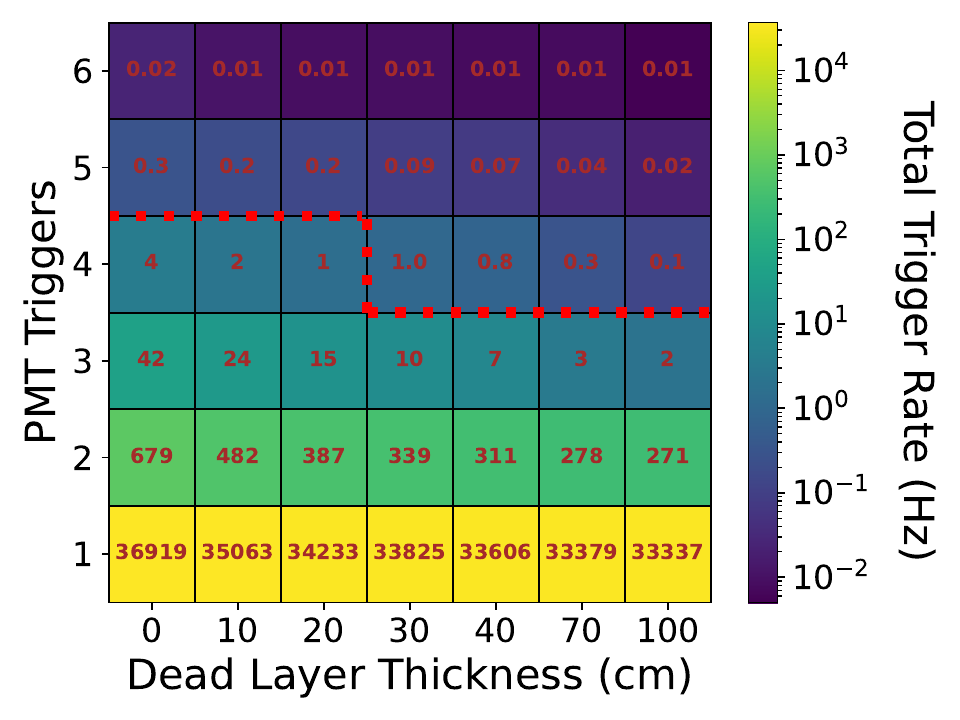} 
    \caption{}
\end{subfigure}
\begin{subfigure}{0.495\textwidth}
    \centering
    \includegraphics[width=\textwidth]{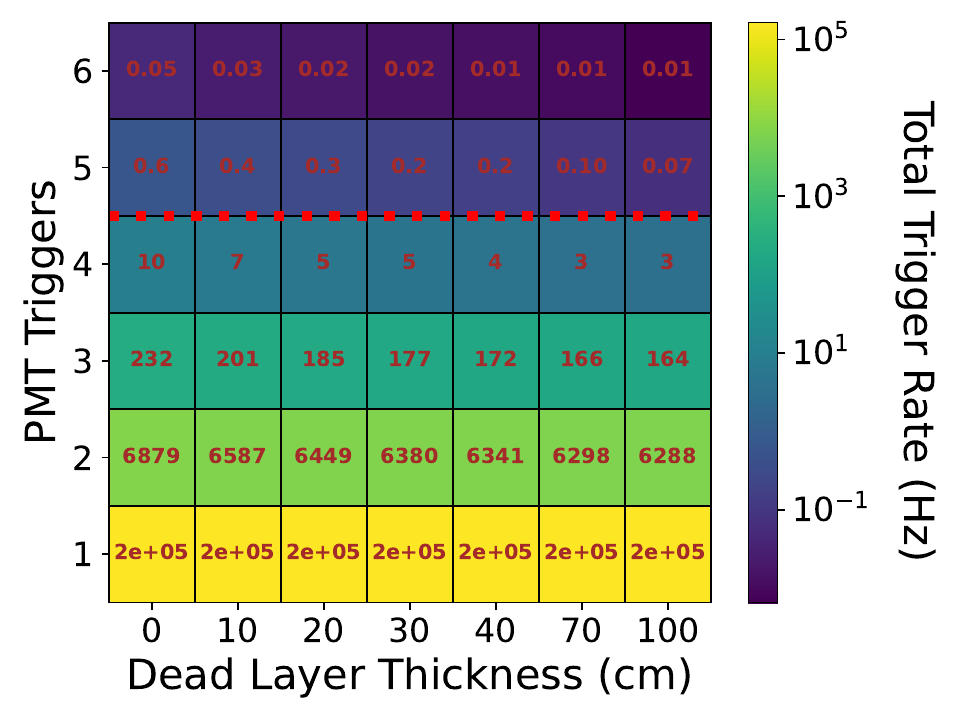}
    \caption{}
\end{subfigure}

\label{}
\centering
\begin{subfigure}{0.495\textwidth}
    \centering
    \includegraphics[width=\textwidth]{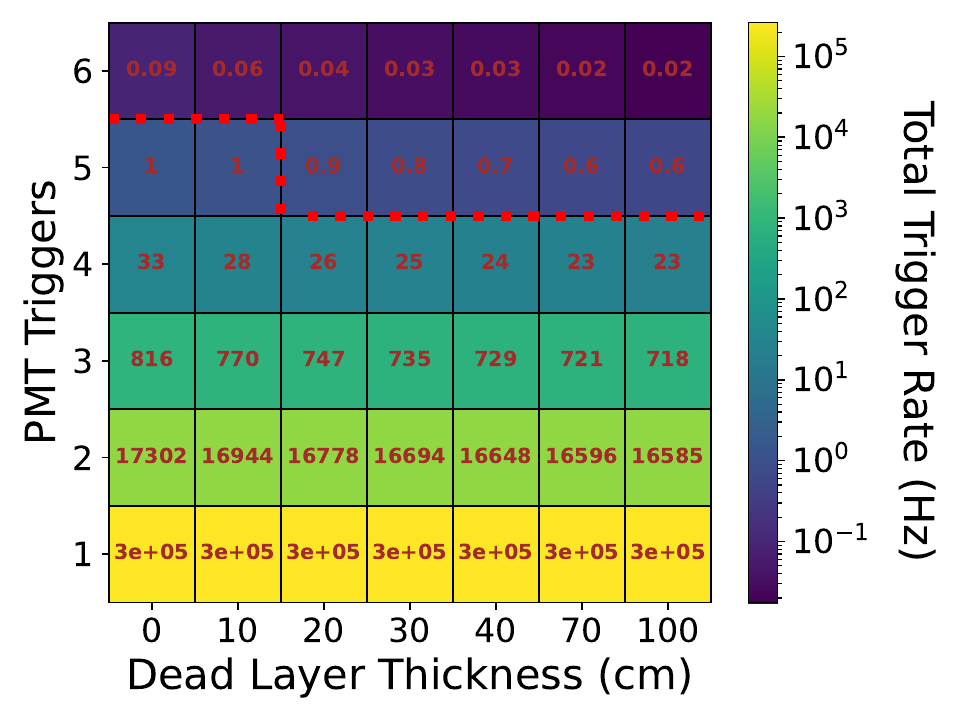}
    \caption{}
\end{subfigure}
\caption{\label{Fig:Total_Rate.pdf} Total background trigger rate for different configurations. Three different dark count rates are used a) 1200~Hz b) 6000~Hz c) 10000~Hz. Above the dashed red line denotes configurations in which the trigger rate is less than 1~Hz. }
\end{figure}

 Thermionic emissions inside the photocathodes of the PMTs can induce triggers that must be folded into the ambient gamma background event rate from Sec.~\ref{subsubsec:Ambient_Gamma_Background_Rate}. By treating the dark counts as a single independent rate they can be added, using combinatorial probabilities, with the ambient gamma rate. This gives the total background rate experienced by the muon veto system. The procedure to calculate the total background rate is outlined in App.~\ref{appendix:bck_rate_calculations}. Based on the results from Fig.~\ref{Fig:Muon_Time_Distribution.pdf} a conservative time window of 500~ns was assumed. Three different dark count rates were considered: $R_d=1200$~Hz which is the rate found by the XENON-1T~\cite{geis2018xenon1t} experiment for the same PMTs, $R_d=6000$~Hz which is the typical dark rate ``quoted" by Hamamatsu and $R_d=10000$~Hz which is the maximal dark count rate quoted by Hamamatsu~\cite{Hamamatsu_PMT_Manual}. The total rate, which includes both dark counts and ambient gamma triggers, is shown in Fig.~\ref{Fig:Total_Rate.pdf} as a function of dead layer and PMT coincidence. The configurations above the dashed red lines represent cases where the total rate is below 1~Hz.

Comparing Fig.~\ref{Fig:Ambient_Gamma_Rate_2D.pdf} to Fig.~\ref{Fig:Total_Rate.pdf} shows that the addition of the dark counts limits the allowed configurations. In the most restrictive case ($R_d=10000$~Hz) the configuration that optimizes the veto efficiency is a 6-fold PMT coincidence requirement and 0~cm dead layer. Again, while contrasting Fig.~\ref{Fig:Dead_Layer_Comparison} and~\ref{Fig:Total_Rate.pdf} it is clear that a higher-fold PMT coincidence will be more effective than an increase in dead layer for reducing the total background rate while maintaining an acceptable efficiency. However, adopting a 30~cm dead layer is a more realistic and practical option as this accounts for the physical length of the PMT and ensures that the required scaffolding used to support the PMT is behind the reflector. The corresponding muon veto efficiency, for the highest dark count rate and 30~cm dead layer, is then {99.58~$\pm$~0.16~$\%$}, and the shower veto efficiency is 41.2~$\pm$~5.6\%. The muon veto efficiency has remained effectively unchanged from the value determined in Sec.~\ref{subsubsec:Ambient_Gamma_Background_Rate} but the shower efficiency has decreased by 15\%. The radiogenic contribution to the PMT background rate from the steel tank is expected to be a few orders of magnitude below the ambient gamma contribution and was not included in this analysis. The count rate is expected to be minimally affected by the Rn contamination in the water, as the level of Rn contamination in COSINUS is expected to be comparable to $10^{-6}$ Bq/kg as reported by Borexino \cite{balata1996water}. This similarity is expected as COSINUS plans to use similar water distillation methods to those used in \cite{balata1996water}. The total radiogenic background recorded from the selected PMT is 7.3 Bq per PMT, with contributions from K-40 at 3.84 Bq per PMT, U-series at 1.89 Bq per PMT, and Th-series at 1.57 Bq per PMT, which is insignificant.

Future work will involve characterizing the experimental dark rate of the PMTs for the COSINUS experiment. 

\subsection{Systematic Uncertainties}
The uncertainties quoted in this manuscript, so far, represent the statistical error found during the Monte Carlo evaluation. This section discusses the systematic uncertainties due to the physics models employed in various code bases. Similar reasoning has been discussed in a previous COSINUS article about the passive shielding simulations~\cite{angloher2022simulation}.

The MUSUN code, which generates the initial energy, position, and direction of the underground muons, was validated against the measured data from the LVD experiment~\cite{kudryavtsev2009muon}. The average energy of muons produced from MUSUN was 273~GeV which agrees with the measured LVD energy of 270 $\pm$ 3 (stat.) $\pm$ 18 (syst.)~GeV. Additionally, Fig. 7 of Ref.~\cite{kudryavtsev2009muon} shows an identical azimuthal angle comparison. The systematic uncertainty of the computed amount, energy spectrum, and angular distribution of muons reaching the LNGS underground halls calculated from MUSUN can be treated as negligible for this study.

There also exist large uncertainties in Geant4 for the yield of muon-induced neutrons. In Ref.~\cite{reichhart2013measurement}, muon-induced neutron yields are compared between measurements and simulations. For an LNGS-like environment the quoted agreement within 25~$\%$ can be an estimate of the systematic uncertainty. For the transport of neutrons ($<$20~MeV) in materials a high-precision, data-driven model is used in Geant4. An agreement of 20$\%$ between Geant4 and MCNPX~\cite{MCNPX} was found~\cite{lemrani2006low} and can be assumed to be an upper limit on the systematic error on low-energy neutron transport. Once the muon veto is operational there will be a dedicated paper comparing the experimental results to the simulations and evaluating the accuracy of these programs for use in underground muon production.

\subsection{Final configuration}\label{subsec:Final_configuration}

Based on the study performed in Sec.~\ref{subsec:Muon_Illumination_Map}-~\ref{subsec:Dead_Layer} the following configuration is recommended:

\begin{itemize}
    \setlength{\itemindent}{1cm}   
    \item 28 PMTs (8-inch diameter R5912-30 Hamamatsu)
    \item Specular reflector with a reflectivity of $>$ 90\%.
   \item Optical dead layer on all sides of the tank (top, bottom, and wall) between 30-40~cm (final size to be decided by engineering considerations) in order to reduce the ambient gamma rate, allow for lower fold PMT coincidence triggering, and account for the practical length of the PMT.
   \item Trigger at a single PE threshold with a greater than 4-fold PMT coincidence (depending on the eventual dark count rate) and a coincidence time window of at least 500~ns.
\end{itemize}

The above conditions, assuming a dark count rate of 1200~Hz, will give a veto efficiency of {99.63~$\pm$~0.16~$\%$} for muon events and 44.4~$\pm$~5.6\% for shower events. As shower events will make up five percent of the total cosmogenic background this gives a total veto efficiency of 97.0~$\pm$~0.3~\%. Therefore, the proposed water Cherenkov detector of the \mbox{COSINUS} experiment will reduce the total cosmogenic neutron background from 3.5~$\pm$~0.7~cts$\cdot$kg$^{-1}$$\cdot$year$^{-1}$ to an acceptable rate 0.11~$\pm$~0.02~cts$\cdot$kg$^{-1}$$\cdot$year$^{-1}$. To provide a model-independent check of the \mbox{DAMA/LIBRA} experimental result an exposure of 1000~kg$\cdot$days is required, with the inclusion of the muon veto, the cosmogenic neutron background will be less than one event during this exposure.

\section{Conclusion}

COSINUS uses NaI crystals as cryogenic calorimeters to search for a direct dark matter signal. An extremely low background rate is required for any successful rare-event search experiment and COSINUS employs both passive and active shielding to achieve this. In this work, a detailed Monte Carlo simulation was performed to design an optimal cosmogenic muon veto. Different experimental configurations, trigger conditions, and reflector types were studied in order to maximize the efficiency of the veto at tagging neutrons produced by cosmic ray muons while maintaining acceptable background trigger rates in the PMTs. With the final configuration presented in Sec.~\ref{subsec:Final_configuration} the obtained veto efficiency in tagging muon and shower events is found to be 
{99.63~$\pm$~0.16~$\%$} and 44.4~$\pm$~5.6\%  respectively which gives a total veto efficiency of $97.0~\pm~0.3\% $. This allows the experiment to reach a muon-induced neutron background of 0.11~$\pm$~0.02~cts$\cdot$kg$^{-1}$$\cdot$year$^{-1}$. For the proposed COSINUS-$1\pi$ exposure of 1000~kg$\cdot$days this will result in less than one background event in the region of interest.

\clearpage
\appendix

\pagestyle{plain}

\section{Background rate calculations}
\label{appendix:bck_rate_calculations}
Important background triggers considered for the COSINUS muon veto are ambient gammas and thermionic emissions inside the PMTs photocathodes (dark counts). Both the arrival of such gammas and the occurrence of dark counts are modeled as Poissonian processes, i.e. the probability $P$ for an event with occurrence rate $R$ to happen within a time interval $\Delta T$ is given by
\begin{equation}
\label{eq:definition_probability_from_rate}
    P(\text{at least one occurrence}) = 1 - P(\text{no occurrence}) =  1-e^{-\Delta T R}.
\end{equation}
The ambient gammas and dark count rates are substantially larger than experimental requirements allow. For the muon veto to achieve the required total background trigger rate of $R_{\text{bck}}<1$ Hz, a higher-fold PMT trigger coincidence has to be demanded (c.f.~Sec.~\ref{subsubsec:Ambient_Gamma_Background_Rate}).\par
The dark counts are assumed to be independent, i.e. the probability $P_{d,L}^{(N)}$ for an $L$-fold coincidence out of $N$ PMTs due to dark counts with rate $R_d$ is described by the binomial distribution ($C$ denotes the binomial coefficient):
\begin{equation}
\label{eq:dark_counts}
    P_{d,L}^{(N)} = C_{N,L}\left[1-e^{-\Delta TR_d}\right]^Le^{-(N-L)\Delta T R_d},\quad C_{N,L}\equiv \frac{N!}{L!(N-L)!}.
\end{equation}
An ambient gamma on the other hand can trigger multiple PMTs and these events are correlated. The rates $R_{\gamma,L}$ at which $L$-fold PMT coincidences happen due to ambient gammas are accessible through simulation. These numbers intrinsically respect all correlations and using Eq.~\eqref{eq:definition_probability_from_rate} the corresponding probabilities $P_{\gamma, L}$ can be derived.\par
Nevertheless, the $R_{\gamma,L}$ were obtained by simulating one photon at a time and without PMT dead-time. Therefore, $L$-fold coincidences due to simultaneously occurring lower-fold coincidences, as well as repeatedly triggering the same PMTs by different photons, are not accounted for. The simulated rates $R_{\gamma,L}$ hence underestimate the true rates for $L$-fold coincidences caused by ambient gammas. The required corrections will be calculated in App.~\ref{appendix:subsection:true_L_fold_coincidence}.\par
Once these corrected rates $\widetilde{R}_{\gamma,L}$ (resp. $\widetilde{P}_{\gamma,L}$) are known, they are combined with the dark counts from Eq.~\eqref{eq:dark_counts} to give the total probability of finding $L$ background PMT triggers in a time window $\Delta T$
\begin{equation}
    \label{eq:cfoldcoincidence}
    P^{L\text{-fold}}_{\text{bck}} = \sum_{k=0}^L \widetilde{P}_{\gamma,k}P_{d,L-k}^{(N-k)}.
\end{equation}
The corresponding $L$-fold background trigger rates of the muon veto are calculated as $R^{L\text{-fold}}_{\text{bck}} = P^{L\text{-fold}}_{\text{bck}}/\Delta T$, and are used to produce the final results shown in Fig.~\ref{Fig:Total_Rate.pdf}.\par
Notice that because the muon veto is blind to the photons' origin, this framework can readily be extended to include other smaller sources of radiation as well (e.g. ambient radiogenic gammas or radioactive contaminants in the water and PMT components).

\subsection{True L-fold coincidence rate of ambient gammas}
\label{appendix:subsection:true_L_fold_coincidence}
The different arrangements to induce an $L$-fold trigger by lower-fold triggers are captured by the \emph{integer partitions} $\mathcal{P}(L)$, describing the possible ways to represent an integer $L$ by smaller, non-overlapping parts (for overlapping ones, see below). The elements of a partition are called \emph{parts} and will be denoted by $p_j$ ($j$-th element of partition $p$). The \emph{length} $\abs{p}$ of a partition is given by the number of its parts. For any partition $p\in \mathcal{P}(L)$ we have a corresponding pre-factor $F^p_L$ which accounts for combinatorics. It can be shown to take the form
\begin{align*}
    F^{p}_L &= \left\{\begin{array}{cc}
            1, &\quad\text{if}~\abs{p} = 1 \\
            \prod_{j=2}^{\abs{p}}\frac{N'_j!}{(N'_j-p_j)!}\frac{(N-p_j)!}{N!} &\quad\text{if}~\abs{p} > 1
        \end{array}\right., 
\end{align*}
where $N'_j\equiv N-\sum_{i=1}^{j-1}p_i$. The probability $P_{\gamma,L}^{\text{unique}}$ for uniquely triggering $L$ out of $N$ PMTs due to ambient gammas is obtained by summing over all partitions $p\in\mathcal{P}(L)$, multiplying $P_{\gamma,1,2,...,L}$ according to their parts, and weighting them with $F^p_L$
\begin{equation*}
    P_{\gamma,L}^{\text{unique}} = \sum_{p \in \mathcal{P}(L)}F^p_LP_{\gamma,1}^{n^p_1}P_{\gamma,2}^{n^p_2}\dots P_{\gamma,L}^{n^p_L},
\end{equation*}
where $n^p_k$ specifies the number of occurrences of the integer $k$ in the partition $p$
.\par
This sum does not include situations where, e.g., two 2-fold triggers combine with a 3-fold trigger due to one of the three PMTs being triggered twice (overlapping parts). Such combinations increase the complexity of the general problem substantially which is why an upper bound was manually calculated up to $L$=6, added to $P_{\gamma,L}^{\text{unique}}$, and used in subsequent calculations.\par
Once an $L$-fold coincidence happens (however it is arranged), an infinite number of additional photons could attempt to trigger these same $L$ PMTs but will not increase the $L$-fold-ness. Including these contributions can be accomplished by a geometric series
\begin{align}
        \widetilde{P}_{\gamma,L} &= P_{\gamma,L}^{\text{unique}}\left[1+\left\{\frac{C_{L,1}}{C_{N,1}}P_{\gamma,1} + \frac{C_{L,2}}{C_{N,2}}P_{\gamma,2} + \dots + \frac{C_{L,L}}{C_{N,L}}P_{\gamma,L}\right\}\right. \nonumber \\
        &\quad\quad\quad\quad\quad\quad\left.+ \left\{\dots\right\}^2 + \left\{\dots\right\}^3 + \dots\right] \nonumber\\ 
        &= P_{\gamma,L}^{\text{unique}}\left[\sum_{k=0}^{\infty}\left\{\sum_{l=1}^L\frac{C_{L,l}}{C_{N,l}}P_{\gamma,l}\right\}^k\right] = \frac{P_{\gamma,L}^{\text{unique}}}{1-\sum_{l=1}^L\frac{C_{L,l}}{C_{N,l}}P_{\gamma,l}}. \label{eq:corrected_gamma_rate}
\end{align}
Notice that here the $P_{\gamma,l}$ appear instead of $P_{\gamma,l}^{\text{unique}}$ to avoid double-counting\footnote{Since all combinations of $l$-fold coincidences are included due to the powers in the geometric series, we naturally account for, e.g., 5-fold coincidences being caused by a 3- and 2-fold coincidence. Notice further, that overlapping configurations are not a problem here.}.\par
Finally, $\widetilde{P}_{\gamma,L}$ and Eq.~\eqref{eq:definition_probability_from_rate} are used to calculate the actual $L$-fold ambient gamma trigger rates, $\widetilde{R}_{\gamma,L}$, which were used to produce Fig.~\ref{Fig:Ambient_Gamma_Rate_2D.pdf} and~\ref{Fig:Total_Rate.pdf}.
\section*{Acknowledgement}
This work has been supported by the Austrian Science Fund FWF-funded stand-alone project AnaCONDa, and the Gran Sasso Science Institute GSSI. We are grateful to INFN and LNGS for their support to COSINUS. 

\bibliographystyle{unsrt}
\bibliography{bibliography}
\end{document}